\documentclass{nature}

\usepackage{graphics,epsf,epsfig,graphicx}

\title{Dense cloud cores revealed by CO in the low metallicity dwarf galaxy WLM}

\author{Monica Rubio$^1$, Bruce G. Elmegreen$^2$,
Deidre A. Hunter$^3$, Elias Brinks$^4$, Juan R.
Cort\'es$^{5,6}$, Phil Cigan$^7$}

\begin{document}

\maketitle

\begin{affiliations}

\item Departamento de Astronom\'ia, Universidad de Chile, Casilla 36-D, Santiago, Chile
\item IBM Research Division, T.J. Watson Research Center, 1101 Kitchawan Road, Yorktown Heights, NY 10598, USA
\item Lowell Observatory, 1400 West Mars Hill Road, Flagstaff, Arizona 86001 USA
\item Centre for Astrophysics Research, University of Hertfordshire, Hatfield AL10 9AB, UK
\item Joint ALMA Observatory, Alonso de C\'ordova 3107, Vitacura, Santiago, Chile
\item National Radio Astronomy Observatory Avenida Nueva Costanera 4091, Vitacura, Santiago, Chile
\item New Mexico Institute of Mining and Technology, Socorro, NM 87801, USA

\end{affiliations}

\begin{abstract}
Understanding stellar birth requires observations of the clouds in which they form.
These clouds are dense and self-gravitating, and in all existing observations, they are
molecular with H$_2$ the dominant species and CO the best available
tracer\cite{mckee07,kennicutt12}. When the abundances of carbon and oxygen are low
compared to hydrogen, and the opacity from dust is also low, as in primeval galaxies and
local dwarf irregular galaxies\cite{remy13}, CO forms slowly and is easily destroyed, so
it is difficult for it to accumulate inside dense clouds\cite{beuther14}. Here we report
interferometric observations of CO clouds in the local group dwarf irregular galaxy
Wolf-Lundmark-Melotte (WLM)\cite{leaman12}, which has a metallicity that is 13\% of the
solar value\cite{lee05,asplund09} and 50\% lower than the previous CO detection
threshold. The clouds are tiny compared to the surrounding atomic and H$_2$ envelopes,
but they have typical densities and column densities for CO clouds in the Milky Way. The
normal CO density explains why star clusters forming in dwarf irregulars have similar
densities to star clusters in giant spiral galaxies. The low cloud masses suggest that
these clusters will also be low mass, unless some galaxy-scale compression occurs, such
as an impact from a cosmic cloud or other galaxy. If the massive metal-poor globular
clusters in the halo of the Milky Way formed in dwarf galaxies, as is commonly believed,
then they were probably triggered by such an impact.
\end{abstract}

WLM is an isolated dwarf galaxy at a distance of $985\pm33$ kpc\cite{leaman12}. Like
other dwarfs, the relative abundance of supernova-processed elements (``metals'') like
Carbon and Oxygen is low\cite{lee05}, $12+\log(O/H)=7.8$, compared to $8.66$ for the
Milky Way\cite{asplund09}.  Low C and O abundances, along with the correspondingly low
abundances of other processed elements and dust, make the CO molecule rare compared to
H$_2$, and this calls into question the standard model of star formation in CO-rich
clouds\cite{mckee07}. In fact, the star formation rate\cite{hunter10} compared to the
existing stellar mass is actually high in WLM: 0.006 $M_\odot$ yr$^{-1}$ of new stars for
a total stellar mass\cite{zhang12} of $1.6\times10^7$ $M_\odot$ is 12 times higher than
in the Milky Way, where the star formation rate\cite{chomiuk11} is
$\sim1.9\pm0.4\;M_\odot$ yr$^{-1}$ and the stellar mass is $6.4\pm0.6\times
10^{10}\;M_\odot$\cite{mcmillan11}. Thus WLM forms stars efficiently even with a
relatively low abundance of CO.

To understand star formation in metal-poor galaxies, which include the most
numerous galaxies in the local universe, the dwarfs, plus all primeaval galaxies,
we previously searched for CO(3-2) in WLM using the APEX telescope
\cite{elmegreen13}, discovering it in two unresolved regions at an abundance
relative to H$_2$ that was half that in the next-lowest metallicity galaxy, the
Small Magellanic Cloud. Now, with the completion of the new millimeter and sub-mm
wavelength interferometer Atacama Large Millimeter Array (ALMA), we have imaged
these two regions in CO(1-0) and resolved the actual molecular structure.

The ALMA maps with $6.2\times4.3$ pc spatial resolution (HPBW), 5 mJy sensitivity, and
0.5 km s$^{-1}$ velocity resolution (FWHM) contain 10 CO clouds with an average radius of
2 parsecs and an average virial mass of $2\times10^3\;M_\odot$. Figure 1 shows the CO
emission with black contours superposed on HI in green and H$\alpha$ in red. The insert
shows a color composite of the optical image in green (V-band), the FUV {\em GALEX} image
in blue and the HI in red. A [CII]$\lambda$158 $\mu$m image from the {\it Herschel} Space
Observatory\cite{pilbratt10} is superposed on the Southeast region in blue\cite{cigan15}.
The [CII] is from a photodissociation region including ionized carbon; it is 5 times
larger in size than the CO core, indicating a gradual transition between low density
atomic gas to high density molecular gas.

Figure 2 shows the contours and spectra of each cloud. The spectral signal-to-noise
averages 10 when smoothed to the typical linewidth of 2 km s$^{-1}$. Velocities for HI
emission are indicated by a bar below each CO spectrum. The cloud properties are
summarized in Table 1. The radii $R$ range from 1.5 to 6 pc, obtained using the equation
$R=(A/\pi)^{0.5}$ for area $A$, with $A$ determined after deconvolution by quadratic
difference with the beam area. The sum of all the line emission measured by ALMA is
within a factor of 2 of the total emission found at $18^{\prime\prime}$ resolution by the
APEX telescope. The linewidths were corrected for instrumental spectral broadening.

Virial masses for the CO clouds were calculated from the relation $M_{\rm
vir}(M_\odot) =  1044 R \sigma^{2}$ for $R$ in pc and Gaussian linewidths $\sigma$
in km s$^{-1}$. The CO luminosity in K km s$^{-1}$ pc$^{2}$ was calculated from
$L_{CO} = 2453 S_{CO} \Delta V D^{2}$ for integrated emission $S$ in Jy km
s$^{-1}$, FWHM of the line $\Delta V$ in km s$^{-1}$, and distance $D$ in Mpc.
Figure 3 shows the relationships between these values including other dwarf
galaxies (all for CO(1-0)). The CO clouds in WLM satisfy the usual correlations
although they are the smallest seen for any of these galaxies. Higher resolution
observations should reveal small clouds and/or cores in other galaxies too, but
the main point is that WLM has no CO clouds as large as those seen elsewhere.

The virial mass gives some perspective on the conversion from CO luminosity to
mass derived previously\cite{elmegreen13}, which was $\alpha_{\rm
CO}\sim124\pm60\;M_\odot\;{\rm pc}^{-2}\left( \rm{K \;km\; s}^{-1}\right)^{-1}$
for the NW region. This value for $\alpha$ was derived from the dust-derived H$_2$
column density. If instead we take the virial masses and CO luminosities in Table
1, we find that the mean ratio is $\alpha_{\rm vir}\sim28\pm28\;M_\odot\;{\rm
pc}^{-2}\left( \rm{K \;km\; s}^{-1}\right)^{-1}$. If the clouds are not
gravitationally bound, then $\alpha_{\rm vir}$ would be smaller. The difference
between these two $\alpha$ values arises because most of the H$_2$ volume has no
CO emission, which apparently exists only in the densest cores of the H$_2$
clouds. For the Milky Way, CO and H$_2$ have about the same extent in star-forming
clouds, making $\alpha_{\rm CO}\sim4\;M_\odot\;{\rm pc}^{-2}\left( \rm{K \;km\;
s}^{-1}\right)^{-1}$. When CO does not fill an H$_2$ cloud, $\alpha$ can be small
for each CO core but large for the total H$_2$ cloud. If the purpose of $\alpha$
is to determine the total H$_2$ mass in a region based on $L_{\rm CO}$, then the
large value should be used.

The self-gravitational boundedness of the CO clouds can be estimated from the
general requirement of an associated H$_2$ density of $\sim10^3$ cm$^{-3}$ for
collisional excitation\cite{glover12a}. In fact, the virial density of the CO
clouds is comparable to this, $n(H_2) = 4.1\times10^{-21}$ g cm$^{-3}$ ($\sim10^3$
cm$^{-3}$), from the ratio of the virial mass ($\sim2\times10^3\;M_\odot$) to the
cloud volume ($4\pi R^3/3$ for $R\sim2$ pc). Thus the clouds could be marginally
bound.

Another measure of CO density is from pressure equilibrium between the CO regions
and the weight of the overlying HI and H$_2$ layers.  The H$_2$ mass column
density, $\Sigma_{\rm H2}$, comes from the difference between the total gas column
density derived from the dust emission and the HI column density observed at 21
cm. For the NW region\cite{elmegreen13}, $\Sigma_{\rm H2}=31\pm15\;M_\odot$
pc$^{-2}$. Adding the HI column density\cite{elmegreen13} gives $\Sigma_{\rm
total}=58\pm15\;M_\odot$ pc$^{-2}$. The corresponding pressure from self-gravity
is $(\pi/2)G\Sigma_{\rm total}^2\sim1.6\times10^{-11}$ dynes. Considering the
typical CO velocity dispersion for our clouds of $\sigma\sim0.9$ km s$^{-1}$, the
ratio of the core pressure to the square of the CO velocity dispersion is the
equilibrium core density, $1.9\times10^{-21}$ g cm$^{-3}$, corresponding to $500$
H$_2$ cm$^{-3}$. Thus the virial density, excitation density, and pressure
equilibrium density are all about $10^3$ cm$^{-3}$.

A condition for molecules in the Milky Way is a threshold extinction of $A_{\rm
V}=0.3$ mag for H$_2$ and $\sim1.5$ mag for CO\cite{glover12b}. These correspond
to mass column densities of $6.1\;M_\odot$ pc$^{-2}$ and $30.3\;M_\odot$ pc$^{-2}$
in the solar neighborhood. In WLM where the metallicity is 13\% solar, the mass
thresholds are $47 \;M_\odot$ pc$^{-2}$ and $230 \;M_\odot$ pc$^{-2}$ for the same
extinctions, respectively.   The first is satisfied by the HI+H$_2$ envelope of
the CO cores ($\sim58\;M_\odot$ pc$^{-2}$) and the second is satisfied by the
total column density of $220\;M_\odot$ pc$^{-2}$ calculated from the HI and H$_2$
envelope, plus the H$_2$ from the embedded CO core itself (as determined from the
CO virial mass, $2\times10^3\;M_\odot$, and ALMA measured radius, $2$ pc). These
results suggest that the CO clouds in WLM are normal in terms of density,
pressure, and column density, which explains why they lie on the standard
correlations in Figure 3. They also appear to be marginally self-bound by gravity,
suggesting they are related to star formation. Their properties are typical for
parsec-size molecular cloud cores in the solar neighborhood\cite{heyer09}.

Our observation explains why star clusters have about the same central densities
in dwarf irregular\cite{billett02} and spiral galaxies\cite{tan13} even though
the ambient gas density in dwarfs is much less than in spirals\cite{eh15}. If the
unifying process for star formation is the need to form CO and other asymmetric
molecules for cooling (however, see\cite{glover12b,krumholz12}), then the
similarity between the CO cores in the two cases accounts for the uniformity of
clusters. The small mass of the CO cores in WLM also explains why most dwarf
galaxies do not form high mass clusters\cite{billett02}.  The CO parts of
interstellar clouds are smaller at lower metallicities, so the clusters that
result are smaller too. For example, there are no massive young clusters in these
regions of WLM\cite{billett02}. This lack of massive clusters is usually
attributed to sparse sampling of the cluster mass distribution function at low
star formation rates\cite{billett02}, but the present observations suggest it
could result from some physical reason too, like the lack of massive CO clouds at
low metallicity.

When the local dwarf galaxies NGC 1569 and NGC 5253 formed massive clusters,
there was a major impact event to increase the pressure and mass at high
density\cite{johnson10,turner15}. Such an impact would also seem to be needed for
the formation of old halo globular clusters, which are massive and low
metallicity like their former dwarf galaxy hosts\cite{bekki08b,elm12}.

\begin{addendum}
\item We wish to thank Phil Massey and the Local Group Survey team for the use of
    their H$\alpha$ image of WLM.  Ms. Lauren Hill made the color composite insert in
    Figure 1. MR would like to thank Cynthia Herrera (NAOJ)
    and Jorge Garcia (JAO, ALMA) for support with the CASA implementation
    to reduce the raw data and A. Rojas for support in the ALMA data reduction. MR is grateful
    to A. Leroy for providing the galaxy
   data to produce Figure 3. MR thanks the ALMA
Director for the invitation to spend her 2015 sabbatical leave at the Joint ALMA
Observatory (JAO) in Santiago, where this article was finished. PC is grateful to Lisa
Young and Suzanne Madden for
    invaluable guidance on Herschel data reduction. MR wishes to acknowledge support
    from CONICYT (CHILE) through FONDECYT grant No. 1140839. MR is partially
    supported by CONICYT project BASAL PFB-06. The contributions from DAH were funded
    by the Lowell Observatory Research Fund. PC acknowledges support from NASA JPL
    RSA grant 1433776 to Lisa Young and grant 1456896 to DAH.
    ALMA is a partnership of ESO (representing its member states), NSF (USA) and NINS
    (Japan), together with NRC (Canada) and NSC and ASIAA (Taiwan), in cooperation with
    the Republic of Chile. The Joint ALMA Observatory is operated by ESO, AUI/NRAO and
    NAOJ.  The National Radio Astronomy Observatory is a facility of the
    National Science Foundation operated under cooperative agreement by Associated
    Universities, Inc.

\item[Author Contributions] DAH, Principle Investigator of the ALMA
    proposal, identified likely CO sources from the re-processed
    data files
    using a direct search for
    significant emission in each frequency channel and for continuous emissions in
    adjacent channels.
    MR re-processed the ALMA results from the originally calibrated
    data delivered by ALMA to get better sensitivity and resolution, finalized the identification
    of emission sources, extracted spectra of the sources, produced Figures
    1 and 2, and produced the measurements in Table 1.
    BGE wrote the text of the manuscript and interpreted the main science results.
    EB oversaw the technical application of radio interferometry
    to molecular line mapping, and determined the noise limitations and deconvolution
    strategy for the angular size and velocity width measurements.
    JRC made the size and
    line-width measurements, produced the virial masses and CO luminosities,
    determined the main observational parameters and made Figure 3.
    PC reduced the Herschel [CII] data and made the [CII] map used in Figure 1.
    All authors contributed to the discussions leading to this manuscript.

\item[Author Information] This paper makes use of the following ALMA data:
    ADS/JAO.ALMA\#2012.1.00208.S. Reprints and permissions information is
    available at www.nature.com/reprints. The authors have no competing financial
    interest in the work described. Correspondence and requests for
    materials should be addressed to: bge@us.ibm.com.

\end{addendum}

\begin{table}
\tiny \caption{Properties of WLM CO clouds}\label{table1}

\begin{tabular}{lcccccccccc}

\hline
Region & RA & Dec & Pk Inten & $V_{\rm LSR}$ & Flux Den. & Radius & $\sigma$ & $M_{\rm vir}$ & $L_{\rm CO}$ \\
& & & (mJy) & (km s$^{-1}$) & (Jy km s$^{-1}$) & (pc) & (km s$^{-1}$) & (M$_\odot$) & (K km s$^{-1}$ pc$^2$)\\
\hline
NW-1	&00 01 57.162 	&-15 27 00.00	&12.2	    &$-131.79\pm0.19$	&$0.0368\pm0.0038$	&$2.21\pm1.11$	&$1.05\pm0.17$	&$2548\pm1522$	&$81.47\pm8.39$\\
NW-2	&00 01 57.291 	&-15 26 52.80	&16.1  	&$-136.42\pm0.18$	&$0.0254\pm0.0026$	&$1.49\pm0.77$	&$0.84\pm0.28$	&$1087\pm919$	&$56.23\pm5.69$\\
NW-3	&00 01 57.901	&-15 26 58.00	&10.8	    &$-126.27\pm0.15$	&$0.048\pm0.0048$	&$2.69\pm1.35$	&$0.75\pm0.14$	&$1561\pm985$	&$106.26\pm10.71$	&\\
NW-4	&00 01 58.079	&-15 27 00.12	&12.2	    &$-125.38\pm0.16$	&$0.0248\pm0.0026$	&$2.69\pm1.35$	&$0.57\pm0.14$	&$898.4\pm637$	&$54.90\pm5.84$\\
SE-1	&00 02 01.485	&-15 27 42.65	&10.8	    &$-121.85\pm0.18$	&$0.0513\pm0.0022$	&$1.68\pm0.87$	&$0.77\pm0.18$	&$1037\pm720$	&$113.57\pm11.46$\\
SE-2	&00 02 01.761	&-15 27 55.83	&13.3	    &$-118.18\pm0.16$	&$0.0212\pm0.0023$	&$<1$	        &$0.61\pm0.23$	&$<390\pm300$        &$46.93\pm5.02$\\
SE-3	&00 02 01.801	&-15 27 51.78	&14.3	    &$-120.00\pm0.12$	&$0.0304\pm0.0031$	&$2.21\pm1.15$	&$0.69\pm0.09$	&$1113\pm653$	&$67.30\pm6.96$\\
SE-4	&00 02 01.864	&-15 28 00.52	&8.77	&$-118.01\pm0.17$	&$0.258\pm0.0026$	&$6.01\pm1.20$	&$1.32\pm0.14$	&$10881\pm3209$	&$571.17\pm57.20$\\
SE-5	&00 02 02.101	&-15 27 58.23	&6.92	&$-117.21\pm0.48$	&$0.0304\pm0.0032$	&$2.02\pm0.96$	&$1.81\pm0.57$	&$6896\pm5426$	&$67.30\pm7.16$\\
SE-6	&00 02 02.222	&-15 27 52.08	&13.7	    &$-117.79\pm0.12$	&$0.0311\pm0.0032$	&$3.37\pm1.06$	&$0.63\pm0.15$	&$1383\pm805$	&$68.85\pm7.11$\\
\hline
\end{tabular}
\end{table}

\clearpage
\begin{figure}
\centering
\includegraphics[width=6.5in]{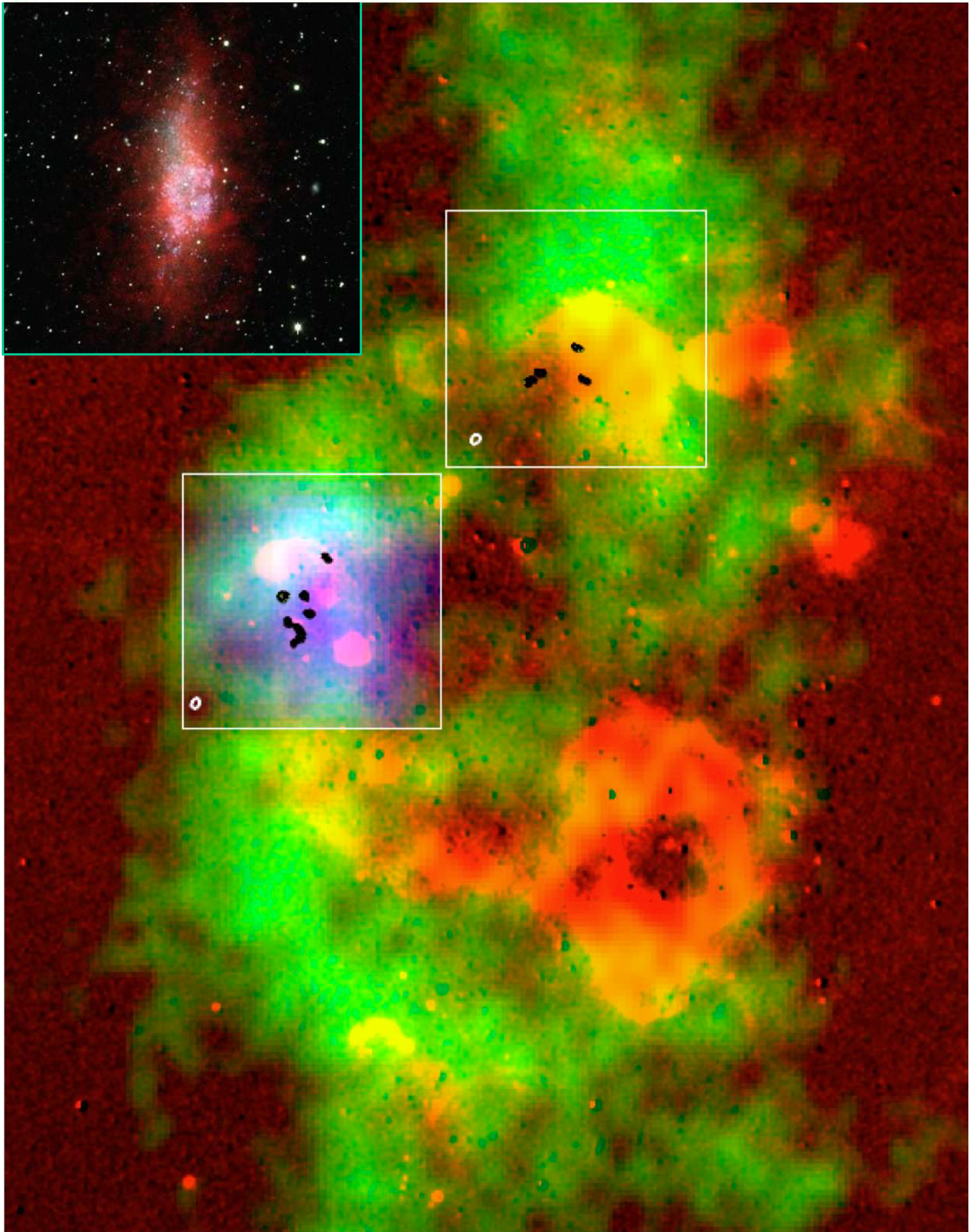}
\caption{Tiny CO clouds in WLM. A color composite of the various gas phases in
WLM: green is the HI \cite{hunter12}, red is H$\alpha$\cite{massey07}, and blue is
[CII]$\lambda$158 $\mu$m\cite{cigan15}. The CO emission is shown as black single
contours inside the 1 arcmin x 1 arcmin white squares that outline the area mapped
in $^{12}$CO (1-0) by ALMA.  The synthesized ALMA beam (0.9"x1.3") is shown in the
lower left corner of each square. The inset in the upper left is the full view of
WLM obtained by combining HI and optical data: red is HI, green is $V$-band, and
blue is {\it GALEX} FUV\cite{hunter12}}.
\end{figure}

\begin{figure}
\centering
\includegraphics[width=6.5in]{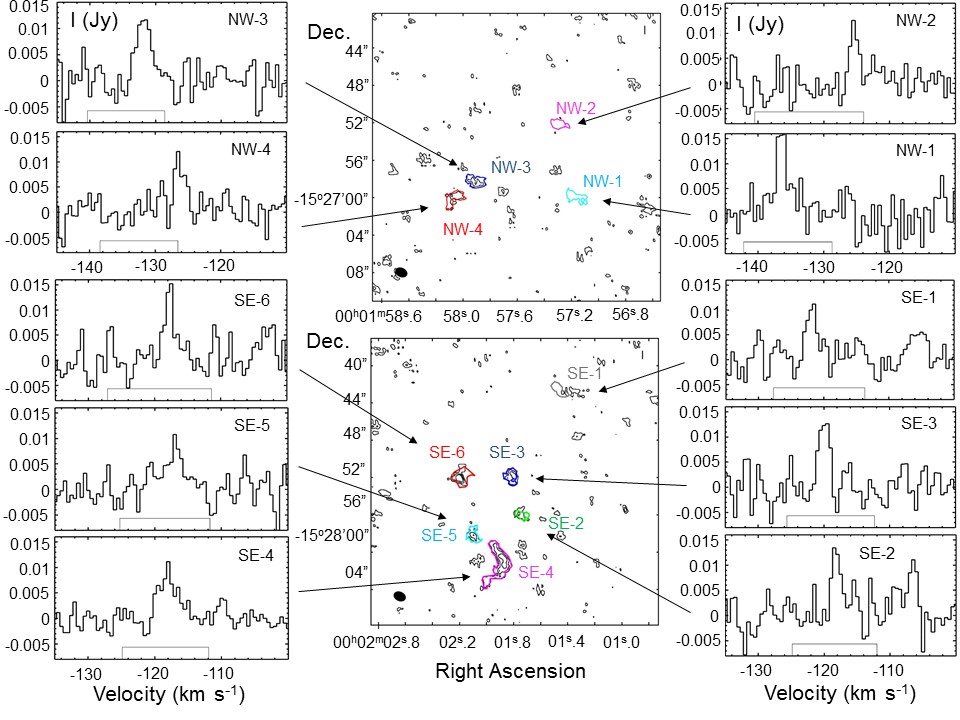}
\caption{CO clouds and spectra. CO contour maps of the integrated emission
starting at the $2-\sigma$ level (RA and DEC in J2000.0 coordinates) . Different
CO clouds are identified by color. The ALMA beam is the black ellipse in the lower
left corner. The CO spectrum corresponding to each detection is plotted. The
velocity for HI emission (FWHM) is shown as a rectangular box on the abscissa
(Local Standard of Rest); the CO velocities agree with the HI.}
\end{figure}

\begin{figure}
\centering
\includegraphics[width=6.5in]{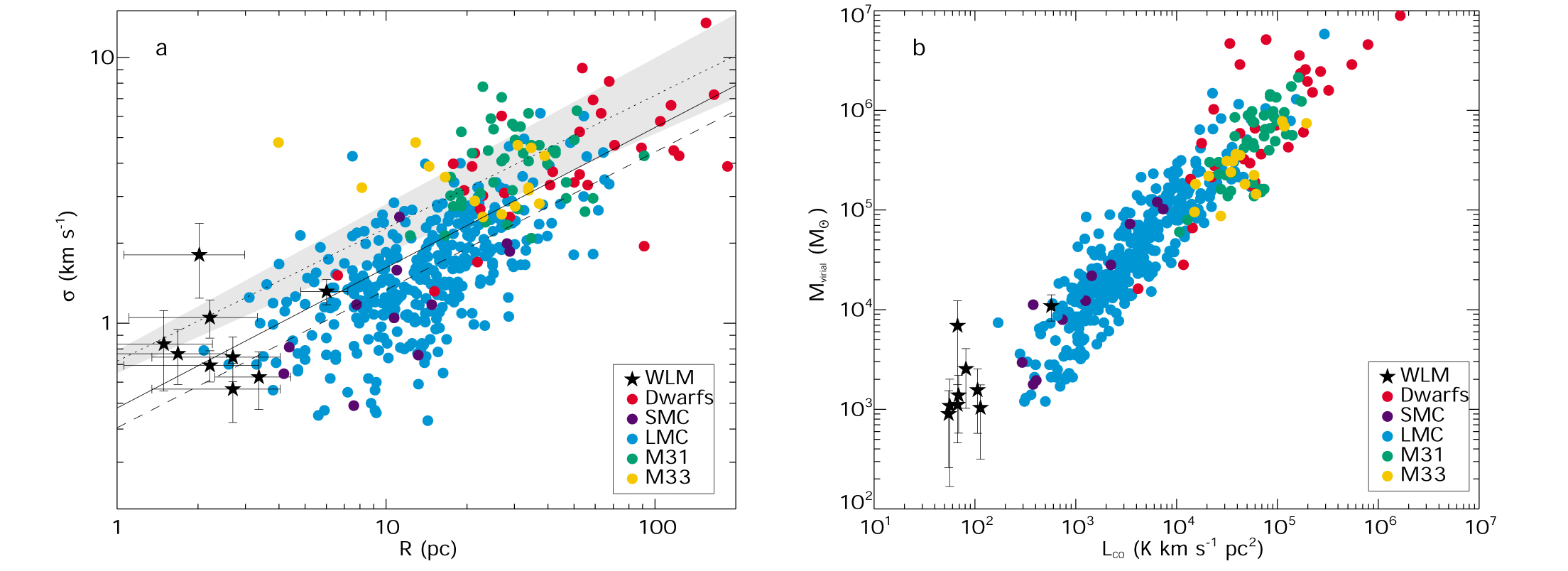}
\caption{Correlations for CO clouds in dwarf galaxies. The symbols refer to
different galaxies (SMC, Dwarfs, M31, and M33\cite{bolatto08}; LMC\cite{wong11}).
(a) CO line width $\sigma$ versus radius $R$; the solid line is a fit to WLM, the
SMC and dwarf galaxies: $\sigma({\rm km}\;{\rm s}^{-1}) = (0.48 \pm 0.08) R({\rm
pc})^{0.53 \pm 0.05}$ and the dashed line includes also the LMC: $\sigma({\rm
km}\;{\rm s}^{-1}) = (0.40 \pm 0.03) R({\rm pc})^{0.52 \pm 0.03}$. The black
short-dashed line and the gray area indicate the standard relation for the Milky
Way\cite{solomon87}: $\sigma = (0.72 \pm 0.07) R^{0.50 \pm 0.05}$. $R$ for WLM is
measured in the same way as for the Milky Way and other galaxies. (b) virial mass
versus CO luminosity.}
\end{figure}


METHODS
\subsection{ALMA Observations}

We observed the $^{12}$CO$(J=1-0)$ transition in two regions in WLM using the
Atacama Large Millimeter/submillimeter Array (ALMA) located on the Chajnantor
Plateau in northern Chile during Cycle 1. Observations were carried out on 2013
July 8 and 2014 April 3. The ALMA receivers were tuned to the ground rotational
transition of Carbon Monoxide, CO(1-0). The interferometer configuration
C32-2/C32-3 provides a maximum baseline of 0.442 km. The observations were done
with a spectral resolution of 122 kHz per channel (0.32 km s$^{-1}$) and total
bandwidth of 468.750 MHz per baseband. The source J2258--2758 was used as a
bandpass calibrator and J2357--1125 was used to calibrate amplitude and phases
with time. To set the absolute flux scale, Uranus was observed. We estimated an
uncertainty in absolute calibration of 10\%.

The data were calibrated, mapped, and cleaned using the ALMA reduction software
CASA (version 4.2.1). Rather than use the pipeline-delivered science data cubes,
we redid the cleaning (i.e., Fourier transform and beam deconvolution) using a
better definition for masking of regions containing emission, and natural
weighting to optimize sensitivity. The maximum angular scale for recovered
emission was estimated to be 15''.

\subsection{Identifying sources}

To make a first cut at identifying sources, we convolved the image cube to a
1.25'' $\times$1.25'' beam and examined a wide velocity range around the velocity
expected from the APEX detection. For the SE region we expected signal around
$V_{\rm LSR}=-120.5$ km s$^{-1}$ and examined $-130.5$ to $-110.5$ km s$^{-1}$.
We detected candidate sources at $-123$ to $-115.5$ km s$^{-1}$. For the NW
region we expected signal around $-130.5$ km s$^{-1}$ and examined $-140.5$ to
$-120.5$ km s$^{-1}$, detecting potential sources at $-139$ to $-121.5$ km
s$^{-1}$. In each velocity channel we looked for knots that had more counts than
the majority of knots that were noise.  Then we looked for signal in nearly the
same location in successive channels, expecting coherence over at least three
channels due to the Hanning-smoothing that had been applied. We also generally
expected the signal to build up and fade away as the channels sampled the source
spectrum. With these criteria, we rated the confidence level of each candidate
source as ``confident'', ``certain'', ``not so certain'', or ``uncertain''. For
the SE region, we identified 9 candidate sources, 6 ranked as ``confident'' or
``certain''. In the NW region, we identified 20 potential sources, 4 ranked as
``certain'' and the rest as less certain.

Based on this identification, we integrated the emission in the velocity range
where CO was seen, and produced the two velocity integrated maps shown in Figure
2 using our reduced new higher sensitivity and velocity resolution cubes. The
velocity resolution of these cubes is 0.5 km s$^{-1}$ per channel. All velocities
are in the Local Standard of Rest (LSR) system. For WLM-SE, 5 integrated maps
were made covering a total LSR velocity range $V_{\rm LSR} = -121$ to $-115.5$ km
s$^{-1}$; the maps spanned velocities of $-121.0$ to $-115.5$, $-121.5$ to
$-119.0$, $-119.0$ to $-117.5$, $-118.5$ to $-117.0$, and $-124.0$ to $-120.5$ km
s$^{-1}$. For WLM-NW, 4 integrated maps were made covering $V_{\rm LSR} = -136.5$
to $-124$ km s$^{-1}$; the individual ranges were $-137.0$ to $-135.5$, $-133.5$
to $-130.0$, $-127.5$ to $-125.5$, and $-127.0$ to $-125.5$ km s$^{-1}$. For
those sources which showed emission at a $3\sigma$ level or above, a spectrum was
obtained integrating over an area delineated by a contour drawn at $2\sigma$ (see
Figure 2) in order not to miss any genuine emission. We also produced
velocity--RA and velocity--Dec maps. Inspecting the CO spectra and the
velocity--position maps, we confirmed 10 CO clouds of the original 20 candidates.
The remaining 10 were deemed of too low signal-to-noise to be included in this
study. On each CO spectrum plot we included the HI emission FWHM velocity width
and converted the HI Heliocentric  to LSR velocity using $V_\mathrm{LSR}$ (i.e.
$V_\mathrm{LSR} = V_\mathrm{Helio}-2.5$ km s$^{-1}$).

The total flux of the 10 clouds resolved with ALMA was compared to the CO(3-2)
flux in our previous APEX observations. We converted the CO(3-2) APEX fluxes from
K km s$^{-1}$ to Jy and assumed a thermal CO(1-0)/ CO(3-2) line ratio of 1. For
WLM-SE we recovered a similar flux of 0.42 Jy in both cases. For  WLM-NW we
measured an ALMA flux of 0.14 Jy while the APEX flux converted to CO(1-0) is 0.66
Jy. The difference in the NW can be due to a different line ratio and thus
different physical conditions, or it could be from weaker emission not included in
our criteria for defining CO clouds, or it could be from emission that is larger
in angular extent than the largest structures measured by the interferometer and
therefore absent from our maps. If we take both regions, then the measured flux
with ALMA is a factor of 2 within the measured flux with APEX.

\end{document}